\documentclass[12p,sort&compress]{elsarticle}
\usepackage{graphicx,epsfig, subfig,tabularx,amscd,amsmath,amssymb,caption,float}
\usepackage{setspace,pbox,makecell}
\journal{Carbon}

\begin{document}
\begin{frontmatter}
\title{A Graphene-Carbon Nanotube Hybrid Material for Photovoltaic Applications} 

\author[egnc,zewail,phys]{Ahmed A. Maarouf\corref{cor1}}
\ead{ahmed.maarouf@egnc.gov.eg}
 
\author[egnc,bue]{Amal Kasry}
\author[ibm]{Bhupesh Chandra}
\author[ibm]{Glenn J. Martyna}

\cortext[cor1]{Corresponding author}

\address[egnc]{Egypt Nanotechnology Center, Cairo University, Giza 12613, Egypt}

\address[zewail]{Center for Fundamental Physics, Zewail City of Science and Technology, Giza 12588, Egypt}

\address[phys]{Department of Physics, Faculty of Science, Cairo University, Giza 12613, Egypt}

\address[bue]{Basic Science Department, Faculty of Engineering, The British University in Egypt, El Sherouk City 11837, Egypt}
\address[ibm]{IBM T. J. Watson Research Center, Yorktown Heights, NY 10598}

\begin{abstract}
 
Large area graphene sheets grown by chemical vapor deposition can potentially be employed as a transparent electrode in photovoltaics if their sheet resistance can be significantly lowered, without any loss in transparency. Here, we report the fabrication of a graphene-conducting-carbon-nanotube (CCNT) hybrid material with a sheet resistance considerably lower than neat graphene, and with the requisite small reduction in transparency. Graphene is deposited on top of a a self-assembled CCNT monolayer which creates parallel conducting paths on the graphene surface. The hybrid thereby circumvents electron scattering due to defects in the graphene sheet, and reduces the sheet resistance by a factor of two. The resistance can be further reduced by chemically doping the hybrid. Moreover, the chemically doped hybrid is more stable than a standalone chemically doped graphene sheet, as the CCNT network enhances the dopant binding. In order to understand the results, we develop a 2D resistance network model in which we couple the CCNT layer to the graphene sheet and demonstrate the model accounts quantitatively for the resistance decrease. Our results show that a graphene-CCNT hybrid system has high potential for use as a transparent electrode with high transparency and low sheet resistance. 
\end{abstract}

%\begin{keyword}
%graphene \sep carbon nanotubes \sep hybrid \sep photovoltaic \sep transparent electrode 
%\end{keyword}

\end{frontmatter}

Graphene, a 2D material consisting of two triangular sub lattices with two carbon atoms per unit cell, has attracted enormous theoretical and experimental interest as a promising material for many applications including electrical devices, chemical separations, and supports for catalysis \cite{neto,lee,agrawal,ang,pan}. It has high mobility \cite{geim2007}, is highly transparent, flexible and relatively chemically stable. These properties make graphene an excellent candidate for use as transparent conducting electrodes (TCE) \cite{pang} in applications such as, solar cell and touch screen technologies. Conventional TCE materials, like indium tin oxide (ITO) \cite{Choi1999}, have the transparency and sheet resistance typically required for TCE applications, but suffer from drawbacks such as the high cost of both the raw materials and their deposition \cite{Chen2001, Leterrier2004,hass2008}. In addition, ITO can not be used for flexible electronics applications.\cite{flexelectronics1, flexelectronics2}  This has created the need for new low-cost TCE materials that exhibit similar or better electrical and optical properties, and has high transparency and low sheet resistance.  A flexible TCE solution is an added bonus.

Pristine undoped graphene has a relatively high sheet resistance, as the Fermi level is at the Dirac point where there are essentially no electronic states to serve as conducting channels. Doping, even by environmental effects, can easily move the Fermi level to a region where the resistance is considerably lower (due to an increase in the density of states). In practice, graphene grown by CVD has grain boundaries formed due to the use of polycrystalline substrates as catalytic material. It also has structural defects and various adsorbed species. The defects and grain boundaries significantly increases the electron scattering in graphene, resulting in its relatively high sheet resistance, even in the presence of doping.\cite{Yu2011} A single layer graphene prepared by CVD typically has a sheet resistance of $1 k\Omega/ \square$ \cite{Atkinson1999}. If the adsorbed dopants are removed by annealing in vacuum, the sheet resistance can reach $5 k\Omega/ \square$, which limits the utilization of graphene in many applications.

Several strategies have been developed to reduce the sheet resistance. One is to stack graphene sheets to form a multilayer structure and thereby introduce more channels for charge transport \cite{Jung2009,Kasry2010}. Another is to dope the multilayers to shift the Fermi level and increase the density of states and thereby the conductivity \cite{Kasry2010,Voggu2008,Lu2009,Eberlein2008}. This approach has been successfully used to reduce sheet resistance to $100 \Omega / \square $  at 80\% transparency\cite{Kasry2010}. Patterning graphene with metal busbar structures via screen printing has also been found to be a successful strategy reducing sheet resistance to $22 \Omega/ \square$ while maintaining high transparency (90\%) limited by contact resistance \cite{Kasry2012}. 

Recently, graphene-nanotubes hybrid structures have attracted theoretical and experimental interest. Pillard nanotubes structures covalently bonded to a graphene sheet have been proposed and fabricated.\cite{pillardhybridtheory1,pillardhybridtheory2,pillardhybridexp1, pillardhybridexp2} Graphene-nanotube hybrid  structures have potential in a wide range of applications, such as nanoelectronics \cite{Kim2014hybrid}, supercapacitors,\cite{Yu2010hybrid, Fan2010hybrid} and chemical applications.\cite{chemapphybrid1, chemapphybrid2}

In this work, to address graphene's relatively high sheet resistance, we connect the grain boundaries in CVD grown graphene with a thin bridge-like conducting material, conducting carbon nanotubes (CCNT) which are ideal for the purpose, to reduce the electron scattering and consequently decrease the sheet resistance. The density of the CCNTs is taken low so that they only connect the grains without reducing graphene's transparency. In order to understand the underlying physics governing the graphene-CCNT hybrid system we construct a physical model. We then present results of experiment. Last, we show results of the numerical computation of the model compared to experiment.

\section{Results and Discussion}

The electronic properties of the hybrid system is primarily dependent on three resistances. The first is that of the graphene sheet itself as imposed by its average grain size\cite{graphenegrainsize,Yu2011}, vacancies, adatoms, and other defects will act as scattering centers for electrons, which will increase the sheet resistance. The second is contact resistance between CCNTs. This resistance depends on the details of the two tubes, including their chiralities, their relative orientation, the crossing angle, and Fermi level of the junction.\cite{crossedtubesexp,crossedtubes} The electronic coupling between a CCNT and a graphene sheet defines the third characteristic resistance of the hybrid system. As in the case of a junction formed from two crossed metallic tubes\cite{crossedtubes}, this coupling is expected to vary with the nanotube chirality and orientation with respect to the underlying graphene lattice. 

To fabricate a graphene-CCNT hybrid film, CCNTs are deposited from a dilute solution on a quartz substrate as explained in Ref[\cite{Tulevski2007,Engel2008}], and a graphene sheet prepared by CVD \cite{Kasry2010} is transferred on to the top of the structure. As a control, a graphene sheet, and two-stacked graphene layers, prepared under the same conditions, are transferred to quartz substrates without CCNT monolayer films. The transmission of the three systems was measured using a UV spectrometer, and the sheet resistance was measured by 4-probe method. 

\begin{figure}[H]
\includegraphics[width=4in]{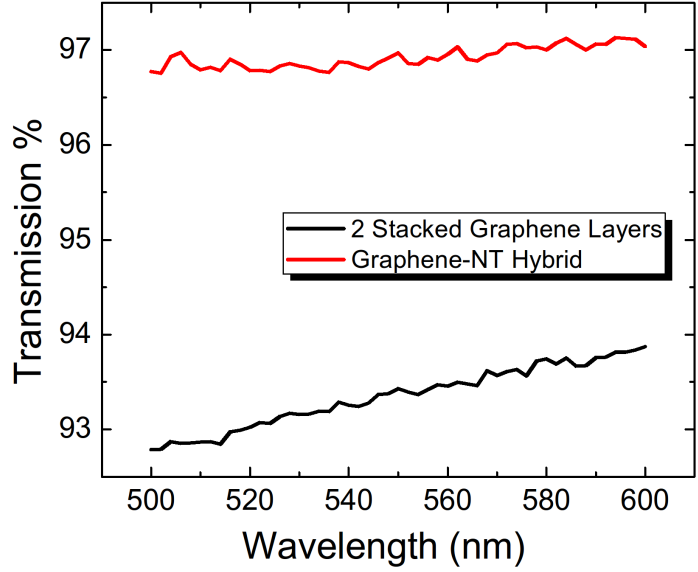}
\caption{The optical density of the different films showing the increase of the sheet resistance of the graphene-CCNT hybrid film at a very high transparency.  }
\label{optical}
\end{figure}

We began the characterization of our first hybrid sample by measuring its optical transparency. Figure \ref{optical} shows the transparency of the hybrid system as a function of wavelength over the visible range, and compares it to the transparency of a bilayer graphene sample. The transparency of the hybrid system is 97\%, higher than that of the graphene bilayer. 

\begin{center}
\begin{figure}[H]
%\def\tabularxcolumn#1{m{#1}}
%\begin{tabularx}{\linewidth}{@{}cXX@{}}
\begin{tabular}{cc}    
   \subfloat[]{\includegraphics[width=2.5in]{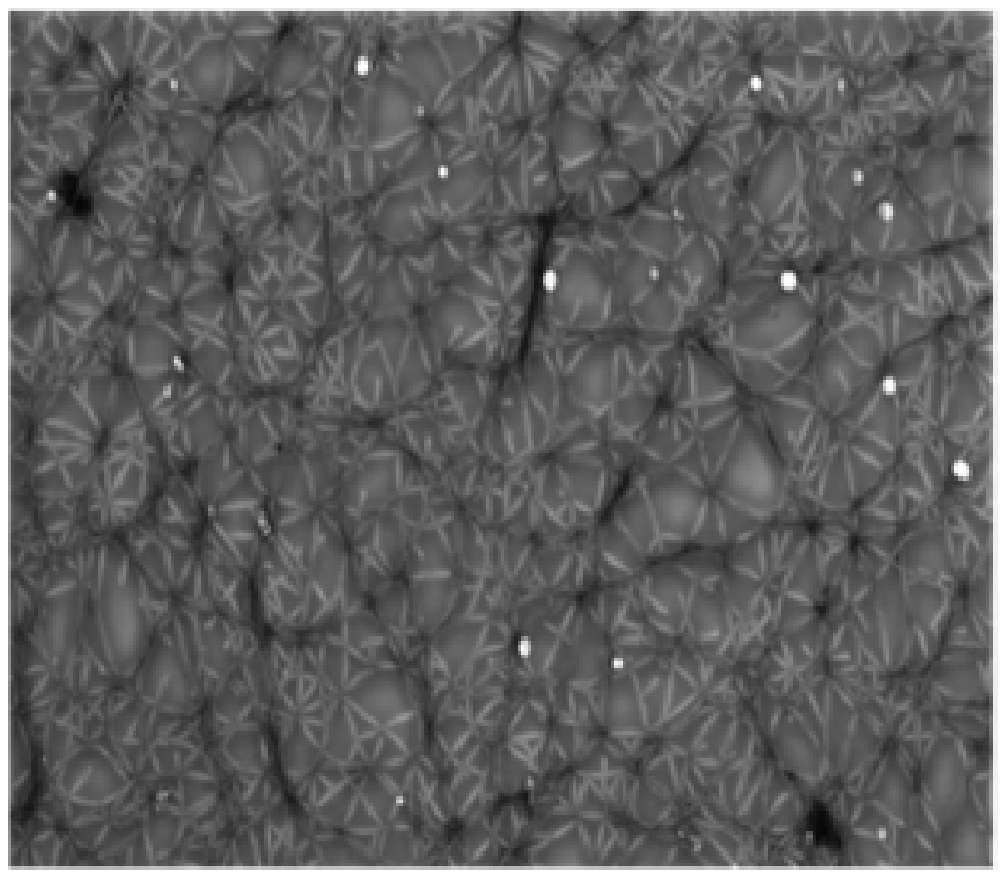}} 
   & \subfloat[]{\includegraphics[width=2.5in]{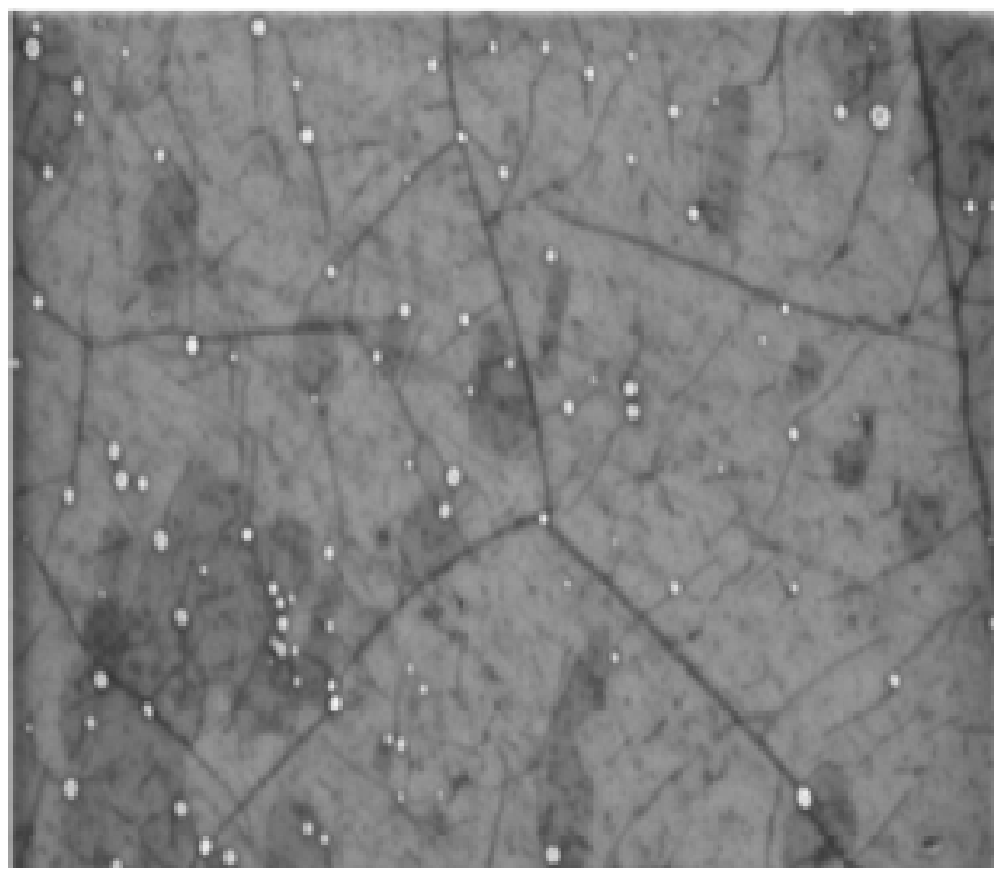}}\\
\end{tabular}
%\end{tabularx}
\caption{(a) SEM image of graphene-CCNT hybrid before annealing. (b) SEM image of graphene-CCNT hybrid after annealing where the nanotubes in the monolayer are not visible any more due to the coupling with the graphene.}\label{sem}
\end{figure}
\end{center}

Various adsorbents existing between the CCNTs and the graphene form an electrically insulating layer that decouples the CCNTs from the graphene. This explains our result that the pristine graphene, the bilayer graphene, and the graphene-CCNT hybrid samples had almost the same resistance. To overcome this problem, the samples were annealed at $600^\circ$C for 10 minutes in vacuum to remove any dopant impurities and improve contact. The resistance of the graphene-CCNT hybrid system is now half that of the annealed graphene, while both exhibited nearly the same transmission. We posit that annealing at this high temperature in vacuum removes undesirable adsorbents and helps the nanotubes achieve good contact with the graphene. This was confirmed by the SEM images of the graphene-CCNT system before annealing ({\bf Fig.\ref{sem}a}). It is clear that CCNTs can be distinguished in the image due to charging. After annealing the nanotubes can not be distinguished and are now coupled to the graphene ({\bf Fig.\ref{sem}b}). The grain boundaries of the graphene layer are clear in both images, and are typically of size $\sim 300$ nm. 

All systems are then doped by immersion in nitric acid solution for 5 minutes as in Ref.\citenum{Kasry2010}. Doping of annealed graphene reduces the sheet resistance from 5.2 $\Omega/ \square$ to 600 $\Omega/ \square$, whereas doping of the annealed hybrid system reduces its resistance from 2.7 $\Omega/ \square$ to 380 $\Omega/ \square$. Doping with nitric acid is known to be unstable, which causes the resistance of samples left in air to increase over time\cite{Kasry2010}. Indeed, leaving our graphene sample in air at room temperature for two days raises its sheet resistance to 780 $\Omega/ \square$. However, the resistance of the hybrid system does not change over the same period. This suggests that the dopants are stabilized by the presence of the CCNT layer, which can be attributed to the trapping of the dopants between the graphene layer and the individual CCNTs {\it along} their length. This interpretation has to be proved by further investigations. The aforementioned results are summarized in table \ref{restable}.  

\begin{center}
\begin{table}[H]
    \begin{tabular}{| l | l | l | l | l |}
    \hline
      & \thead{Graphene} & \thead{Bilayer \\ graphene} & \thead{Hybrid \\ sample 1} & \thead{Hybrid \\ sample 2} \\ \hline
    As prepared & 1.3 $k\Omega$ & 1.2 $k\Omega$ & 1.32 $k\Omega$ & 1.2 $k\Omega$ \\ \hline
    After vacuum annealing & 5.2 $k\Omega$ & 4.1 $k\Omega$ & 2.7 $k\Omega$ & 2.7 $k\Omega$ \\ \hline
    After nitric acid doping & 600 $\Omega$ & 380 $\Omega$ & 367 $\Omega$ & 390 $\Omega$ \\ \hline
    After de-doping & 780 $\Omega$ & 660 $\Omega$ & 380 $\Omega$ & 380 $\Omega$ \\ \hline
    \end{tabular}
    \caption{Sheet resistance values for graphene, bilayer graphene, and the hybrid graphene-CCNT system. The resistance of the hybrid system shows a large degree of stability compared to graphene.}
    \label{restable}  
    \end{table}
\end{center} 

\begin{figure}[H]
\includegraphics[width=5in]{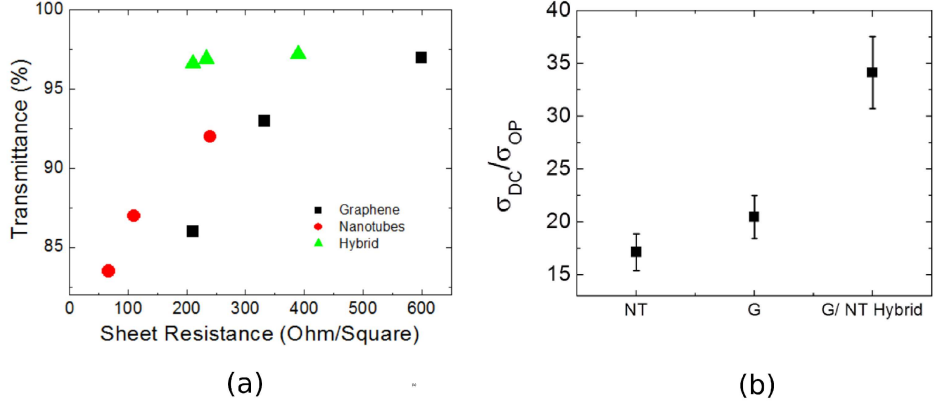}
\caption{(a) Transmission vs sheet resistnace. (b) The optical density of the different films showing the increase of the sheet resistance of the graphene-CCNT hybrid film at a very high transparency.  }
\label{resistance_transmission}
\end{figure}

To study the effect of the CCNT density on the sheet resistance of the hybrid material, we prepared samples with different CCNT monolayer densities. In Fig. \ref{resistance_transmission}a we show our transmission and sheet resistance results for hybrid samples with different CCNT densities, multilayer graphene, and {\it thick} carbon CCNT films. All samples were first annealed then doped by nitric acid as before. All hybrid samples have the same transparency as single layer graphene, confirming their monolayer nature. The sheet resistance of the lowest-density CCNT hybrid sample is about half that of single layer graphene. Increasing the monolayer density 3-fold brings the sheet resistance down to about one third that of single layer graphene. Further increase in the CCNT monolayer density has a minor effect on the sheet resistance of the hybrid. By comparison, similar resistance reduction is achieved with a thick CCNT film with transparency of 90\%, and few layers graphene with a transparency of 85\%. 

The ratio of DC to optical conductivity ($\sigma_{DC}/\sigma_{Op}$) is an important parameter that can be used to evaluate the performance of transparent electrodes. This ratio can be extracted from the measurement of the optical transparency: 
\begin{equation}
T(\lambda) = \left( 1 + \frac{188.5 \Omega}{R_S} \frac{ \sigma_{Op}(\lambda)}{\sigma_{DC}} \right) ^{-2}
\end{equation}
where $\sigma_{DC}$ and $\sigma_{Op}$ are the electrical and optical conductivities respectively, $R_S$ is the sheet resistance and $\lambda$ is the wavelength which the transmission is measured at (550 nm). This equation was used to calculate and compare the ratios between the electrical and optical conductivities for CCNT film, graphene layer, and graphene-CCNT hybrid. Figure \ref{resistance_transmission}b shows this ratio for the three layers where the hybrid has $\sigma_{DC}/\sigma_{Op}= 34$, very close to the ITO value of Ref. \citenum{Choi1999}.

\begin{figure}[H]
\includegraphics[width=5in]{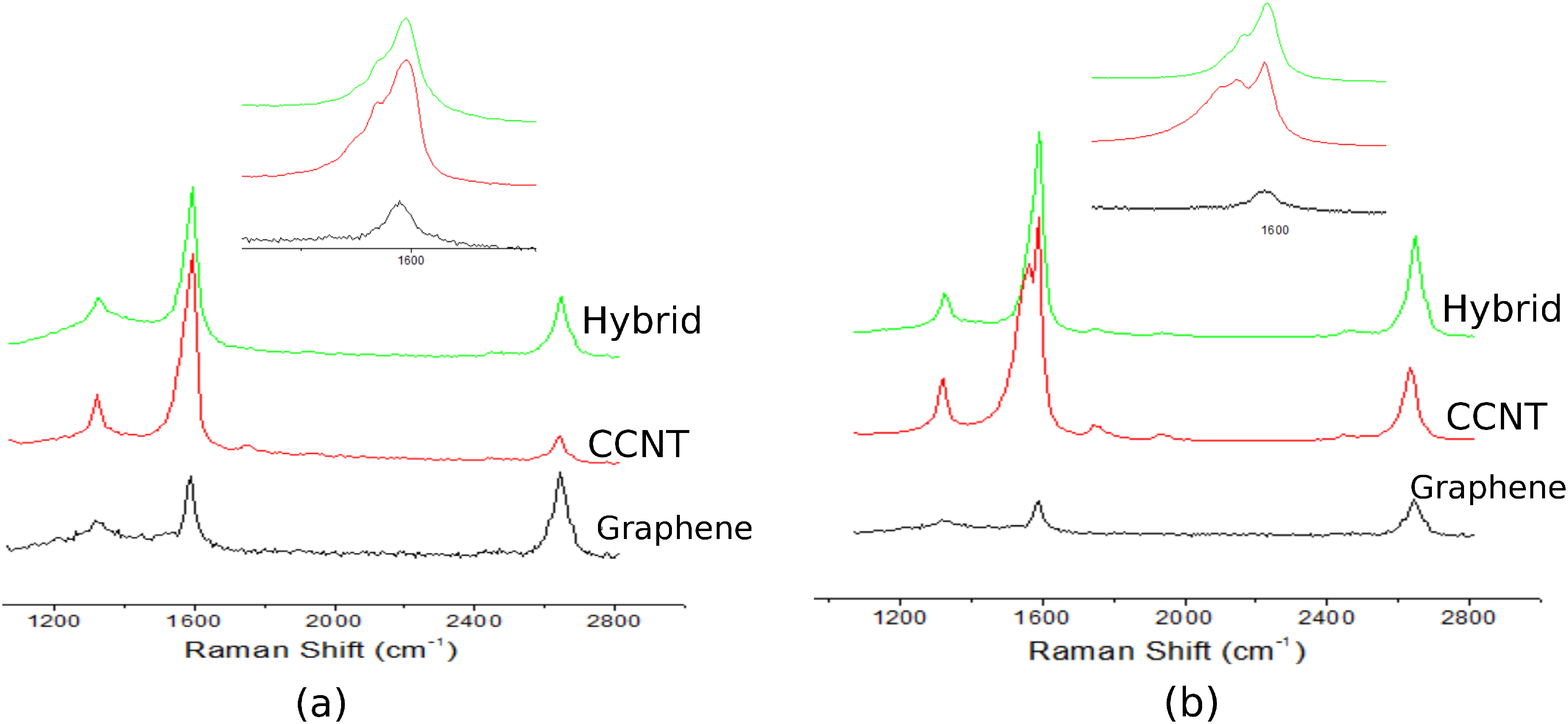}
\caption{ Raman scattering spectra of graphene monolayer, CCNT monolayer, and the hybrid. (a) Sample with lowest density CCNT monolayer. (b) Sample with highest density CCNT monolayer. Insets show the spectra at the G-peak region.}
\label{raman}
\end{figure}

Raman measurements on the hybrid samples were performed, and the results are shown in Fig.\ref{raman}. As we see, the signal of the hybrid is much closer to that of the single layer graphene than to that of the isolated CNNT monolayer (Fig. \ref{raman}a). As the CCNT monolyar density is increased, the Raman CCNT signature starts to appear, as shown in Fig.\ref{raman}b. The insets focuses on the G-peak region.

% Theory

We now explain our model. As a first approximation, and since the nanotubes are randomly dispersed on the graphene sheet, the coupling between the nanotubes and the underlying graphene is taken to be constant - a mean field type approximation. Furthermore, all nanotubes are taken to be conducting, which is justified by the fact that the graphene-nanotube system is chemically doped, although purification methods are being developed to separate conducting from insulating tubes.\cite{georgeali} Since we are interested in the low nanotube-density limit where the nanotube-network percolation itself is absent or weak, contributions from the nanotubes to the graphene conductance will arise from the nanotubes crossing graphene scattering boundaries (e.g. a nanotube crossing over a grain boundary). That is, a nanotube creates another conductance path between neighboring regions, and thus one can think of the nanotube as a local modification of the resistance of the graphene sheet. We further assume that the contact resistance between crossing CCNTs is much higher than that between a CCNT and graphene.\cite{crossedtubesexp,crossedtubes} This is confirmed by the observed significant decrease of the hybrid resistance for CCNT film densities that are well below the CCNT-network percolation limit. 

\begin{center}
\begin{figure}[H]
%\def\tabularxcolumn#1{m{#1}}
%\begin{tabularx}{\linewidth}{@{}cXX@{}}
\begin{tabular}{cc}  
 
   \subfloat[]{\includegraphics[width=2.in]{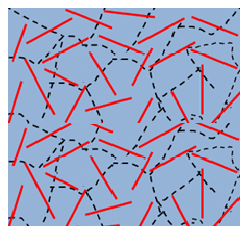}} 
   & \subfloat[]{\includegraphics[width=2.in]{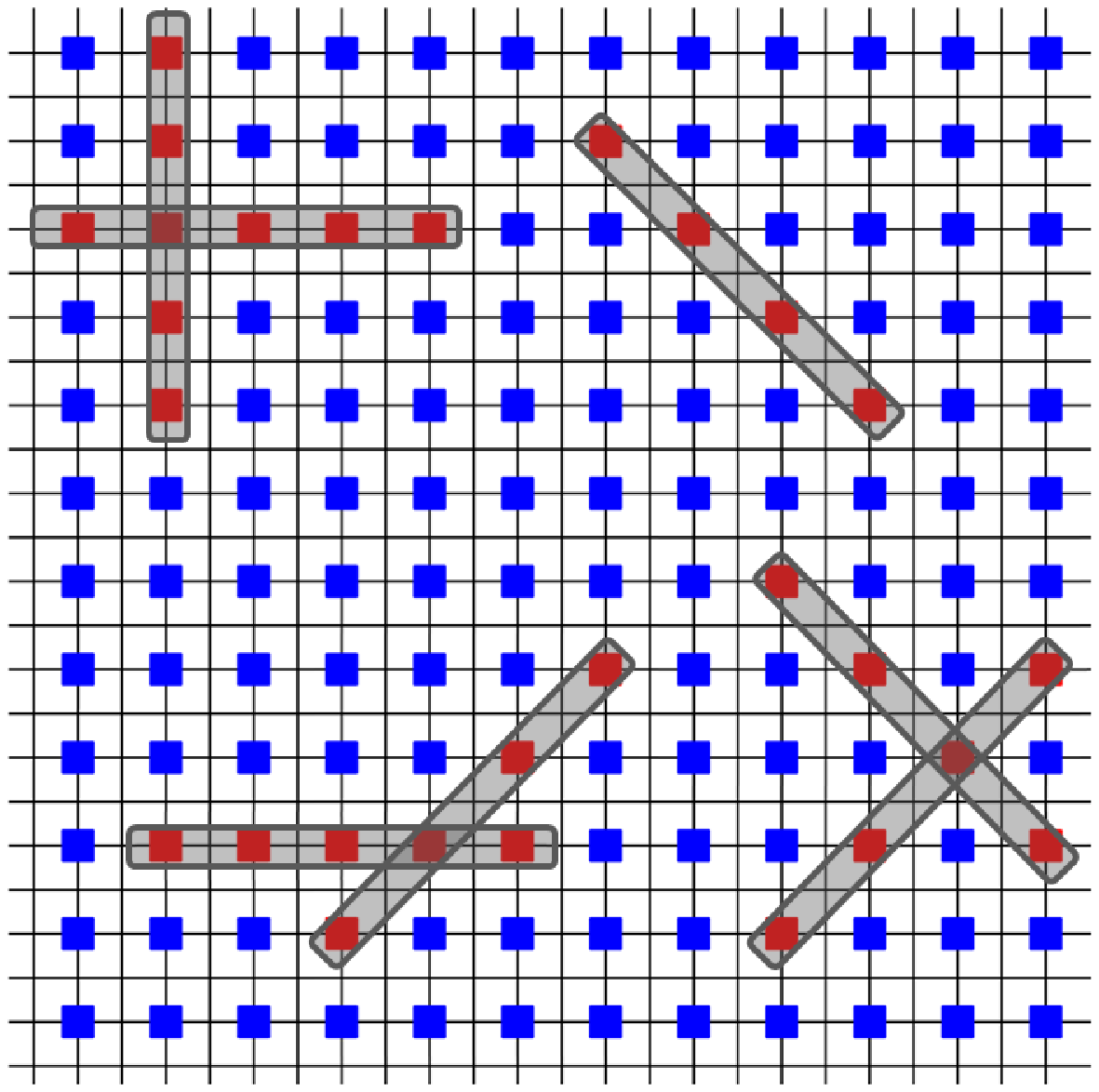}}\\
   
\end{tabular}
%\end{tabularx}
\caption{(a) Schematic showing graphene sheet with a layer of nanotubes. The dotted lines mark the graphene grains and the CCNTs are shown as red lines. (b) A 2D resistance network model of the graphene sheet. The CCNTs increase the sheet conductance by locally changing the model resistance.}\label{model}
\end{figure}
\end{center}

The graphene sheet is treated as a continuum resistive medium, which we discretize into a square grid of tiles. The tile edge, $a_G$, is taken to be smaller than the smallest length in the system (the nanotube length $L_T$). Each tile is connected to its four nearest neighbors using a resistance $r_G$, determined by the experimentally measured graphene sheet resistance. The graphene-CCNT hybrid system is constructed by adding the nanotubes to the graphene sheet at random positions and with random orientations. We model this by adding local parallel resistance $r_{GT}$ when a  randomly placed tube crosses two graphene tiles. Nanotubes are taken to be of uniform size, $L_T \sim 350$ nm, as inferred from SEM images of the physical system. Figure \ref{model} a,b present a schematic of a graphene sheet with grains and the nanotubes acting as bridges connecting the grains. Having understood the basic physics through the model construction, we present the experimental realization of the system.

We next return to theory to interpret the data. The sheet resistance of the modelled hybrid system is determined using a 4-probe resistance calculation. Current is injected from the outer probes, and the voltage drop on the inner two probes is calculated using Kirchhoff's laws. The net node current $I_i$ at node $i$ is related to various node voltages $V_j$ by:
\begin{equation}
\sum_{j=1}^{N} {}^{'}  g_{ij}(V_i - V_j) = I_i,
\end{equation}
where $i = 1,2, .. N$, and the sum excludes the term with $i=j$. The element $g_{ij}$ is the conductance between nodes $i$ and $j$ in the network. In matrix form, Eq. \ref{networkequations} can be written as
\begin{equation}
G V = I,
\label{networkequations}
\end{equation}
where $G$ is called the Kirchhoff's matrix of the network and it is defined as:
\[
G = \left(
\begin{matrix}
g_{1} &  -g_{12}  & \ldots & -g_{1N}\\
-g_{12}  &  g_{2} & \ldots & -g_{2N}\\
\vdots & \vdots & \ddots & \vdots\\
-g_{1N}  &   -g_{2N}       &\ldots & g_{N}
\end{matrix}
\right),
\]
The diagonal terms are defined by current conservation (flux in equals flux out)
\begin{equation}
g_i = \sum_{j=1}^{N} {}^{'}  g_{ij},
\nonumber
\end{equation}
A current vector $I$ containing the net current at each node has zero elements except for the two nodes, $a$ and $b$, in contact with the current carrying probes. Equation \ref{networkequations} is solved for the voltage at the nodes $c$ and $d$ in contact with the two inner probes. The sheet resistance is thus:
\begin{equation}
R = \frac{V_c - V_d}{I},
\end{equation}
where $I$ is the current entering node $a$ and leaving node $b$. The reduction in the sheet resistance of the hybrid system can be described using the fractional decrease in the resistance caused by the CCNT film, defined as:
\begin{equation}
\eta=\frac{R_G - R_H}{R_G},
\end{equation}
where $R_G$ is the sheet resistance of graphene, and $R_H$ is the sheet resistance of the hybrid system.

A sample size of side $L_s =$ 40 microns was simulated. The tile size is taken to be $a_G =$ 0.25 microns matching a typical graphene grain size.\cite{graphenegrainsize,Yu2011} The tile size is smaller than the nanotube length $L_T$, so that each CCNT is guaranteed to bridge at least two tiles, thereby simulating the coupled graphene CCNT system. Reassuringly, further decrease in the tile size $a_G$ does not lead to significant change in the calculated sheet resistance. The resistance parameter $r_G$ is set by fitting the calculated neat graphene sheet resistance to the measured value. The CCNT density, $\rho_{2D}$, and length, $L_T =$ 0.35 microns, are inferred from the SEM images of Fig.\ref{sem}c. The inter-probe distance, $d_{4p}$, of the 4-probe calculation is taken to be 4 microns, so that it spans many graphene-CCNT grains. This way, we have $a_G < L_T << d_{4p} << L_s$. This order secures a quantitatively correct description of the system. A detailed study of the dependence of the calculated sheet resistance of the sample on various parameters is presented in the supplementary information section.

In Fig. \ref{etafig}, we plot the experimental values of $\eta$ for the annealed and doped hybrid system with the results of our calculation. The resistance parameter $r_{GT}$ describing the graphene-CCNT coupling is obtained by fitting model to experiment as described above. For low CCNT densities, the model predicts a linear dependence of $\eta$  on the density. As the density increases above the percolation threshold, not all CCNTs are in contact with the graphene, and since the CCNT-CCNT resistance is much higher than that of the CCNT-graphene, the fractional decrease in the resistance will reach a constant value, which is observed in our experimental data. Such 3D effects are not included in our simple model, which predicts a further increase in $\eta$ at high CCNT densities, where some CCNTs lie on each other rather than on the graphene sheet. The inclusion of such effects would decrease the resistance gain obtained by the CCNT layer, but its treatment would be rather subjective in our phenomenological model. For example, it would have to involve how the CCNTs bend around each other, and how the bend is affected by details such as the number of underlying or overlying CCNTs, as well as the crossing angles. In addition, a model describing how this would affect the CCNT resistance would have to be constructed. However, these details do not change the overall picture captured by the current treatment, but does explain the deviation of our model from our experimental data at high CCNT densities. 

\begin{figure}[H]
\includegraphics[width=4in]{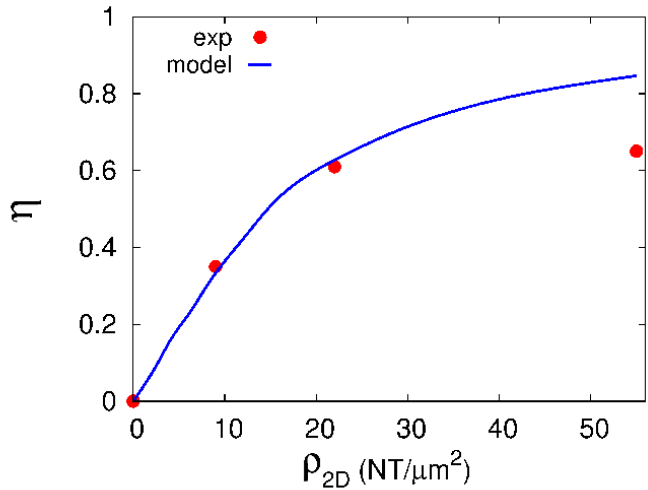}
\caption{A plot of $\eta$ vs the CCNT 2D density $\rho_{2D}$. The model predicts a linear dependence on the CCNT density at lower densities below or at weak percolation of the CCNT network. For higher densities, not all CCNTs contribute to lowering the graphene resistance, $\eta$ saturates. The deviation between the model and the experiment at high density is because 3D effects are not considered in the model.}
\label{etafig}
\end{figure}

%The doped graphene-CCNT system shows much higher doping stability than graphene. Doping of annealed graphene reduces the sheet resistance from 5.2 $k\Omega$ to 600 $\Omega$, whereas doping of the annealed hybrid system reduces its resistance from 2.7 $k\Omega$ to 380 $\Omega$. Doping with nitric acid is known to be unstable, which causes the resistance of samples left in air to their values before doping\cite{Kasry2010}. Indeed, leaving our graphene samples in air for two days raises its sheet resistance to 780 $\Omega$. However, the resistance of the hybrid system does not change. This suggests that the dopants are stabilized by the presence of the CCNT layer, which has to be proved by further investigations. 

\section{Conclusion}
In conclusion, we have shown that a hybrid material made from CCNTs atop 	graphene sheet significantly reduces the sheet resistance compared to neat graphene, with a negligible decrease in transparency. The CCNT film reduces the system sheet resistance by providing alternative current paths which bridge over different scattering regions. The differential reduction in resistance is maximal at low CCNT densities, where the CCNT network has weak or no percolation, indicating that the alternative current paths created are due to bypassing the graphene domain edges. A simple resistance network model successfully explains the experimental behaviour. Furthermore, CCNT network increases the dopant binding to the graphene sheet, thereby enhancing the doping stability. This hybrid graphene-CCNT system has potential applications in various TCE applications. We can combine the CCNT-graphene hybrid with the micro-busbar array of Ref.\citenum{Kasry2012} to obtain a very effective TCE with many technological applications.

\section{Methods}

Graphene films were prepared by the CVD method \cite{Kasry2010}. Graphene was grown on Cu foils; a piece of Cu foil (25 µm thick, Sigma-Aldrich) was placed in a 1-inch diameter quartz furnace tube at low pressure of 60 mTorr. The system was flushed with 6 sccm forming gas (5\% H$_2$ in Ar) for 2 hours at a pressure of 500 mTorr to remove any residual oxygen and water from the system. The concentration of the oxygen and water in the chamber was monitored with a mass spectrometer. The Cu foil was then heated to 975$^\circ$C in forming gas (6 sccm, 500 mTorr) and kept at this temperature for 10 minutes to reduce native CuO and increase the Cu grain size. After reduction, the Cu foil was exposed to ethylene (6 sccm, 500 mTorr) at the same temperature for 10 minutes. The sample was cooled down in forming gas (6 sccm, 500 mTorr) before removing out of the system. PMMA was spin-coated on the graphene layer formed on the Cu foil, and the Cu foil was then dissolved in 1 M iron chloride. The graphene film supported by the PMMA layer was washed with DI water and transferred to a quartz substrate where the PMMA was dissolved in acetone for one hour. The samples were annealed in vacuum at 600$^\circ$C for 10 minutes to burn any remaining PMMA and to get rid of all the dopants.

SWNTs (95\% purity) were purchased from APS-100F, Hanwha Nanotech Corp. A diluted solution was prepared according to REF\cite{Moshammer2009}. The substrates were modified with a 1\% solution of 3-aminopropyltriethoxysilane (APTES) in DI water, to form a monolayer that can act as an adhesion layer between the nanotubes and the substrate. The nanotubes film was deposited by immersing the substrates in the diluted solution for 30 min, and then washing with copious amount of water. The graphene sheets were transferred to the self-assembled nanotubes as explained above. The graphene sheets were transferred to the self-assembled nanotubes as was explained above. The transmission spectrum was measured using a Perkin Elmer Lambda 950 UV/Vis spectrometer. The sheet resistance was measured using a manual four point probe equipment from Signatone, probe distance 1.5 mm.

\section{Supporting Information}

\subsection{Theoretical Model}

We treat the graphene sheet as a continuum resistive medium, which we model by discretizing it using a square grid of tiles. The tile size is taken to be smaller than the smallest length in the problem (the nanotube length $L_T$). Each tile is connected to its four nearest neighbors using a resistance $r_G$, determined by the value of the pristine graphene resistance from our experiments. The tile size, $a_G$,  taken in our simulations was about 250 nm ( i.e. less than the CCNT length). This guarantees that the limit In that sense, the graphene sheet is modeled as a square 2D network of resistors of resistance $r_G$ each. Therefore, the different lengths in the systems are such that $a_G < L_T << d_{4p} < L_S$.  We quantify this statement below.

Resistance networks of various lattice symmetries have been studied before using various analytical and computational approaches. \cite{Atkinson1999,Wu2004,Cserti2011} The focus of previous work was on the two point resistance, which is the resistance between two grid points on the resistance network (usually separated by a few network lattice constants). Here, we model the 4-probe resistance of our graphene-CCNT hybrid system. We choose 4 points across the length of the simulated sample, inject and collect current from the two outer points, and calculate the potential difference across the two inner ones. To calculate the 4-probe resistance, we use Kirchhoff's laws as described in the text.

The simulated sample size was 40 microns $\times$ 40 microns. Our study shows that the 4-probe calculation is independent of the inter-probe distance if it is less than ~ 1/10 of the sample size. Our system has four lengths. In decreasing order, these are: the length defined by the sample dimensions, the inter-probe distance of the four-probe device, the nanotube length, and the tile size of the graphene sheet. This order implies a qualitatively correct description of the system, but to obtain quantitative credibility, one has to explore the dependence of the calculated resistance on the various parameters of the sample in order to get a highly accurate description given the large computational cost involved.

\begin{figure}[H]
\includegraphics[width=4in]{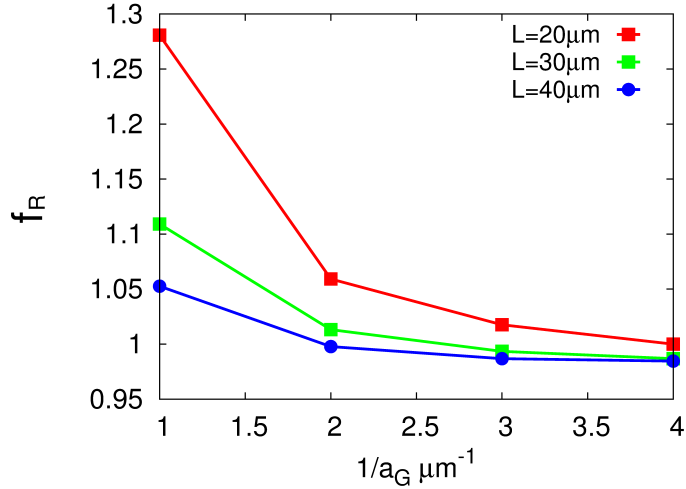}
\caption{A plot of $f_R$ vs the inverse of the tile size $a_G$ for various sample lengths $L_s$.}
\label{supfig1}
\end{figure}

Let us first consider the tile size which we use in our discretization of the graphene sheet. For a given sample dimensions, as the tile size decreases, the description gets closer to the infinite resistances network limit, and the sheet resistance reaches a constant value. This clear in Fig.\ref{supfig1}, where we plot the relative sheet resistance $f_R$ vs the inverse of the graphene sheet grid size, $a_G$. The relative sheet resistance is the sheet resistance normalized by the resistance at some tile size value (chosen to be 1 micron). As we see, a tile size value of 0.3 microns or less is enough to simulate the infinite system limit for sample dimensions of 30 or more microns, with an accuracy better than 1\%.  

\begin{figure}[H]
\includegraphics[width=4in]{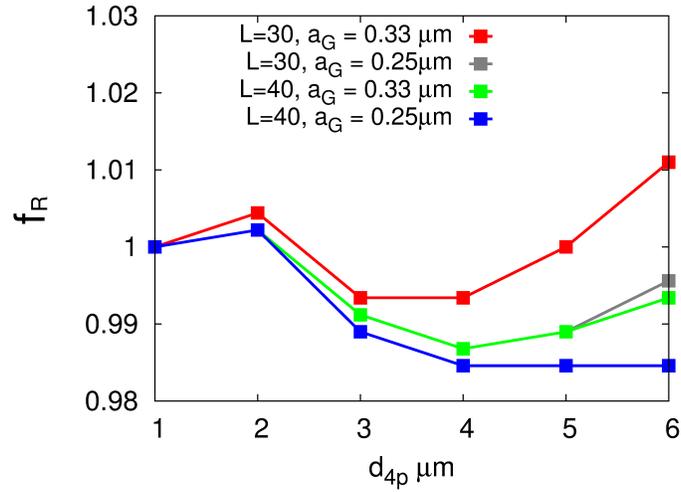}
\caption{A plot of $f_R$ vs the 4-probe measurement inter-probe distance $d_{4p}$ at different tile size $a_G$ and sample length $L_s$.}
\label{supfig2}
\end{figure}

To study the dependence on the four probe distance $d_{4p}$, we plot the relative resistance vs $d_{4p}$ for various samples lengths and tile sizes. Figure \ref{supfig2} shows such dependence, and it is clear that converged results (with difference of less than 1\%) are obtained for $d_{4p}$ = 3 or 4 microns, for tile sizes smaller than 0.3 microns, and sample length 30 microns or higher. Here, a calculation using a smaller 4-probe distance will suffer from finite size effects. 

\begin{figure}[H]
\includegraphics[width=4in]{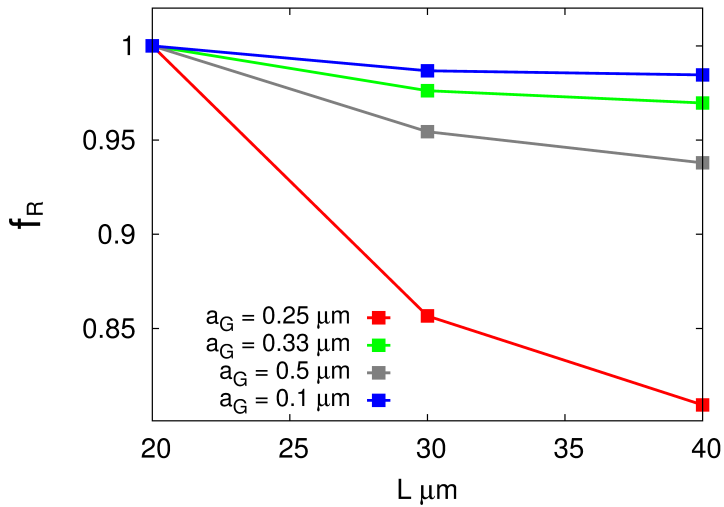}
\caption{A plot of $f_R$ vs sample length $L_s$ at various tile size $a_G$. }
\label{supfig3}
\end{figure}

In fig. \ref{supfig3} we plot the relative resistance $f_R$ vs the square sample length $L$, for various tile sizes, and for a 4-probe distance of 4 microns. As we see, for sample lengths 30 microns or higher, change in the relative resistance is less than 1\%.

Therefore, we choose our simulation parameters as $L_s = 40$ microns, $d_{4p} = 4$ microns, and $a_G = 0.25$ microns. It should be noted here that our SEM images indicate that our CCNTs have an average length of 0.35 microns. Smaller grid sizes enhance the accuracy by a fraction of a percent, but with a huge computational cost.

In our simulations, the CCNTs are generated with random position and orientation on the graphene sheet. A CCNT will span at least two graphene grid squares. At each CCNT density, at least 500 different systems are generated, and their sheet resistances are determined. An average is then obtained.

\bibliographystyle{elsarticle-num}
\bibliography{hybrid}

\begin{thebibliography}{10}
\expandafter\ifx\csname url\endcsname\relax
  \def\url#1{\texttt{#1}}\fi
\expandafter\ifx\csname urlprefix\endcsname\relax\def\urlprefix{URL }\fi
\expandafter\ifx\csname href\endcsname\relax
  \def\href#1#2{#2} \def\path#1{#1}\fi

\bibitem{neto}
A.~H. Castro~Neto, F.~Guinea, N.~M.~R. Peres, K.~S. Novoselov, A.~K. Geim, The
  electronic properties of graphene, Rev. Mod. Phys. 81 (2009) 109--162.

\bibitem{lee}
J.~H. Lee, D.~W. Shin, V.~G. Makotchenko, A.~S. Nazarov, V.~E. Fedorov, Y.~H.
  Kim, J.-Y. Choi, J.~M. Kim, J.-B. Yoo, One-step exfoliation synthesis of
  easily soluble graphite and transparent conducting graphene sheets, Advanced
  Materials 21~(43) (2009) 4383--4387.

\bibitem{agrawal}
J.~Wu, M.~Agrawal, H.~A. Becerril, Z.~Bao, Z.~Liu, Y.~Chen, P.~Peumans, Organic
  light-emitting diodes on solution-processed graphene transparent electrodes,
  ACS Nano 4~(1) (2010) 43--48, pMID: 19902961.
\newblock \href {http://arxiv.org/abs/http://dx.doi.org/10.1021/nn900728d}
  {\path{arXiv:http://dx.doi.org/10.1021/nn900728d}}.

\bibitem{ang}
P.~K. Ang, S.~Wang, Q.~Bao, J.~T.~L. Thong, K.~P. Loh, High-throughput
  synthesis of graphene by intercalation−exfoliation of graphite oxide and
  study of ionic screening in graphene transistor, ACS Nano 3~(11) (2009)
  3587--3594, pMID: 19788171.
\newblock \href {http://arxiv.org/abs/http://dx.doi.org/10.1021/nn901111s}
  {\path{arXiv:http://dx.doi.org/10.1021/nn901111s}}.

\bibitem{pan}
Y.~Pan, H.~Zhang, D.~Shi, J.~Sun, S.~Du, F.~Liu, H.-j. Gao, Highly ordered,
  millimeter-scale, continuous, single-crystalline graphene monolayer formed on
  ru (0001), Advanced Materials 21~(27) (2009) 2777--2780.

\bibitem{geim2007}
a.~K. Geim, K.~S. Novoselov, {The rise of graphene.}, Nature materials 6~(3)
  (2007) 183--91.

\bibitem{pang}
S.~Pang, H.~N. Tsao, X.~Feng, K.~Müllen, Patterned graphene electrodes from
  solution-processed graphite oxide films for organic field-effect transistors,
  Advanced Materials 21~(34) (2009) 3488--3491.

\bibitem{Choi1999}
K.~H. Choi, J.~Y. Kim, Y.~S. Lee, H.~J. Kim, {ITO / Ag / ITO multilayer ® lms
  for the application of a very low resistance transparent electrode}, Thin
  Solid Films 341 (1999) 152--155.

\bibitem{Chen2001}
Z.~Chen, B.~Cotterell, W.~Wang, E.~Guenther, S.-j. Chua, {A mechanical
  assessment of flexible optoelectronic devices}, Thin Solid Films (2001)
  202--206.

\bibitem{Leterrier2004}
Y.~Leterrier, L.~Medico, F.~Demarco, U.~Betz, M.~F. Escola, M.~K. Olsson,
  F.~Atamny, {Mechanical integrity of transparent conductive oxide films for
  flexible polymer-based displays}, Thin Solid Films 460 (2004) 156--166.

\bibitem{hass2008}
J.~Hass, W.~A. de~Heer, E.~H. Conrad, The growth and morphology of epitaxial
  multilayer graphene, Journal of Physics: Condensed Matter 20~(32) (2008)
  323202.

\bibitem{flexelectronics1}
B.~Chandra, H.~Park, A.~Maarouf, G.~J. Martyna, G.~S. Tulevski, {Carbon
  nanotube thin film transistors on flexible substrates}, Applied Physics
  Letters 99~(7) (2011) 072110.

\bibitem{flexelectronics2}
Y.~Liu, H.~Zhou, R.~Cheng, W.~Yu, Y.~Huang, X.~Duan, {Highly flexible
  electronics from scalable vertical thin film transistors.}, Nano letters
  14~(3) (2014) 1413--8.

\bibitem{Yu2011}
Q.~Yu, L.~A. Jauregui, W.~Wu, R.~Colby, J.~Tian, Z.~Su, H.~Cao, Z.~Liu,
  D.~Pandey, D.~Wei, T.~F. Chung, P.~Peng, N.~P. Guisinger, E.~A. Stach,
  J.~Bao, S.-S. Pei, Y.~P. Chen, {Control and characterization of individual
  grains and grain boundaries in graphene grown by chemical vapour
  deposition.}, Nature materials 10~(6) (2011) 443--9.

\bibitem{Atkinson1999}
D.~Atkinson, {Infinite resistive lattices}, American Journal of Physics 67~(6)
  (1999) 486.

\bibitem{Jung2009}
N.~Jung, N.~Kim, S.~Jockusch, N.~J. Turro, P.~Kim, L.~Brus, {Charge transfer
  chemical doping of few layer graphenes: charge distribution and band gap
  formation.}, Nano letters 9~(12) (2009) 4133--7.

\bibitem{Kasry2010}
A.~Kasry, M.~A. Kuroda, G.~J. Martyna, G.~S. Tulevski, A.~A. Bol, {Chemical
  doping of large-area stacked graphene films for use as transparent,
  conducting electrodes.}, ACS nano 4~(7) (2010) 3839--44.

\bibitem{Voggu2008}
R.~Voggu, B.~Das, C.~S. Rout, C.~N.~R. Rao, {Effects of charge transfer
  interaction of graphene with electron donor and acceptor molecules examined
  using Raman spectroscopy and cognate techniques}, Journal of Physics:
  Condensed Matter 20~(47) (2008) 472204.

\bibitem{Lu2009}
Y.~H. Lu, W.~Chen, Y.~P. Feng, P.~M. He, {Tuning the electronic structure of
  graphene by an organic molecule.}, The journal of physical chemistry. B
  113~(1) (2009) 2--5.

\bibitem{Eberlein2008}
T.~Eberlein, R.~Jones, J.~Goss, P.~Briddon, {Doping of graphene: Density
  functional calculations of charge transfer between GaAs and carbon
  nanostructures}, Physical Review B 78~(4) (2008) 045403.

\bibitem{Kasry2012}
A.~Kasry, M.~{El Ashry}, R.~a. Nistor, A.~a. Bol, G.~S. Tulevski, G.~J.
  Martyna, D.~M. Newns, {High performance metal microstructure for carbon-based
  transparent conducting electrodes}, Thin Solid Films 520~(15) (2012)
  4827--4830.

\bibitem{pillardhybridtheory1}
G.~K. Dimitrakakis, E.~Tylianakis, G.~E. Froudakis, {Pillared graphene: a new
  3-D network nanostructure for enhanced hydrogen storage.}, Nano letters
  8~(10) (2008) 3166--70.

\bibitem{pillardhybridtheory2}
F.~D. Novaes, R.~Rurali, P.~Ordej\'{o}n, {Electronic transport between graphene
  layers covalently connected by carbon nanotubes}, ACS Nano 4~(12) (2010)
  7596--7602.

\bibitem{pillardhybridexp1}
R.~K. Paul, M.~Ghazinejad, M.~Penchev, J.~Lin, M.~Ozkan, C.~S. Ozkan,
  {Synthesis of a pillared graphene nanostructure: A counterpart of
  three-dimensional carbon architectures}, Small 6 (2010) 2309--2313.

\bibitem{pillardhybridexp2}
D.~H. Lee, J.~E. Kim, T.~H. Han, W.~J. Hwang, S.~W. Jeon, S.~Y. Choi, S.~H.
  Hong, W.~J. Lee, R.~S. Ruoff, S.~O. Kim, {Versatile carbon hybrid films
  composed of vertical carbon nanotubes crown on mechanically compliant
  graphene films}, Advanced Materials 22 (2010) 1247--1252.

\bibitem{Kim2014hybrid}
S.~H. Kim, W.~Song, M.~W. Jung, M.~a. Kang, K.~Kim, S.~J. Chang, S.~S. Lee,
  J.~Lim, J.~Hwang, S.~Myung, K.~S. An, {Carbon nanotube and graphene hybrid
  thin film for transparent electrodes and field effect transistors}, Advanced
  Materials 26 (2014) 4247--4252.

\bibitem{Yu2010hybrid}
D.~Yu, L.~Dai, {Self-Assembled Graphene/Carbon Nanotube Hybrid Films for
  Supercapacitors}, The Journal of Physical Chemistry Letters 1~(2) (2010)
  467--470.

\bibitem{Fan2010hybrid}
Z.~Fan, J.~Yan, L.~Zhi, Q.~Zhang, T.~Wei, J.~Feng, M.~Zhang, W.~Qian, F.~Wei,
  {A three-dimensional carbon nanotube/graphene sandwich and its application as
  electrode in supercapacitors}, Advanced Materials 22 (2010) 3723--3728.

\bibitem{chemapphybrid1}
X.~Dong, Y.~Ma, G.~Zhu, Y.~Huang, J.~Wang, M.~B. Chan-Park, L.~Wang, W.~Huang,
  P.~Chen, {Synthesis of graphene–carbon nanotube hybrid foam and its use as
  a novel three-dimensional electrode for electrochemical sensing}, Journal of
  Materials Chemistry 22 (2012) 17044.

\bibitem{chemapphybrid2}
G.~K. Dimitrakakis, E.~Tylianakis, G.~E. Froudakis, {Pillared graphene: a new
  3-D network nanostructure for enhanced hydrogen storage.}, Nano letters
  8~(10) (2008) 3166--70.

\bibitem{graphenegrainsize}
P.~Y. Huang, C.~S. Ruiz-Vargas, A.~M. van~der Zande, W.~S. Whitney, M.~P.
  Levendorf, J.~W. Kevek, S.~Garg, J.~S. Alden, C.~J. Hustedt, Y.~Zhu, J.~Park,
  P.~L. McEuen, D.~a. Muller, {Grains and grain boundaries in single-layer
  graphene atomic patchwork quilts.}, Nature 469 (2011) 389--392.
\newblock \href {http://arxiv.org/abs/1009.4714} {\path{arXiv:1009.4714}}.

\bibitem{crossedtubesexp}
M.~S. Fuhrer, J.~Nygård, L.~Shih, M.~Forero, Y.-G. Yoon, M.~S.~C. Mazzoni,
  H.~J. Choi, J.~Ihm, S.~G. Louie, A.~Zettl, P.~L. McEuen, Crossed nanotube
  junctions, Science 288~(5465) (2000) 494--497.

\bibitem{crossedtubes}
A.~A. Maarouf, E.~J. Mele, Low-energy coherent transport in metallic carbon
  nanotube junctions, Phys. Rev. B 83 (2011) 045402.

\bibitem{Tulevski2007}
G.~S. Tulevski, J.~Hannon, A.~Afzali, Z.~Chen, P.~Avouris, C.~R. Kagan,
  {Chemically assisted directed assembly of carbon nanotubes for the
  fabrication of large-scale device arrays.}, Journal of the American Chemical
  Society 129~(39) (2007) 11964--8.

\bibitem{Engel2008}
M.~Engel, J.~P. Small, M.~Steiner, M.~Freitag, A.~A. Green, M.~C. Hersam,
  P.~Avouris, {Thin film nanotube transistors based on self-assembled, aligned,
  semiconducting carbon nanotube arrays.}, ACS nano 2~(12) (2008) 2445--52.

\bibitem{georgeali}
G.~S. Tulevski, A.~D. Franklin, A.~Afzali, {High purity isolation and
  quantification of semiconducting carbon nanotubes via column
  chromatography.}, ACS nano 7~(4) (2013) 2971--6.

\bibitem{Moshammer2009}
K.~Moshammer, F.~Hennrich, M.~M. Kappes, {Selective suspension in aqueous
  sodium dodecyl sulfate according to electronic structure type allows simple
  separation of metallic from semiconducting single-walled carbon nanotubes},
  Nano Research 2 (2009) 599--606.

\bibitem{Wu2004}
F.~Y. Wu, {Theory of resistor networks: the two-point resistance}, Journal of
  Physics A: Mathematical and General 37~(26) (2004) 6653--6673.

\bibitem{Cserti2011}
J.~Cserti, G.~Sz\'{e}chenyi, G.~D\'{a}vid, {Uniform tiling with electrical
  resistors}, Journal of Physics A: Mathematical and Theoretical 44~(21) (2011)
  215201.

\end{thebibliography}

\end{document}